\documentclass[12pt]{article}
\usepackage[margin=2.5cm]{geometry}
\usepackage{amsmath}
\usepackage{amssymb}
\usepackage[utf8]{inputenc}
\usepackage{bbold}
\usepackage{xcolor}
\usepackage[makeroom]{cancel}
\usepackage{mathtools}
\usepackage[colorlinks, allcolors=cyan]{hyperref}
\usepackage[backend=biber, style=phys, sorting=none, citestyle=numeric-comp, url=false, eprint=false]{biblatex}
\appto{\bibsetup}{\sloppy}
\AtEveryBibitem{\clearfield{title}}
\bibliography{bibliography}

\begin{document}

\title{Unification of Conformal and Fuzzy Gravities with Internal Interactions based on the SO(10) GUT \footnote{This article is based on an invited plenary talk given by GZ in ``11th Mathematical Physics Meeting'', 2-6 September 2024, in Belgrade, Serbia.}}

\author{Danai Roumelioti$^1$, Stelios Stefas$^1$, George Zoupanos$^{1,2,3,4,5}$}

\maketitle
\begin{center}
\itshape$^1$Physics Department, National Technical University, Athens, Greece\\
\itshape$^2$ Max-Planck Institut f\"ur Physik, M\"unchen, Germany\\
\itshape$^3$ Universit\"at Hamburg, Hamburg, Germany\\
\itshape$^4$ Deutsches Elektronen-Synchrotron DESY, Hamburg, Germany\\
\itshape$^5$ Arnold-Sommerfeld-Center for Theoretical Physics, LMU, Munich, Germany
\end{center}

\begin{center}
\emph{E-mails: \href{mailto:danai\_roumelioti@mail.ntua.gr}{danai\_roumelioti@mail.ntua.gr}, \href{mailto:dstefas@mail.ntua.gr}{dstefas@mail.ntua.gr}, \href{mailto:George.Zoupanos@cern.ch}{George.Zoupanos@cern.ch}}
\end{center}

\begin{abstract}
We present a unification scenario of the conformal and fuzzy gravities with the internal interactions, based on the observation that the tangent space of a curved space and the space itself do not necessarily have the same dimensions. Accordingly, the construction is based on enlarging the tangent space, while it is formulated in a gauge-theoretical way.
\end{abstract}

\section{Introduction}\label{intro}

The aim of constructing a unified scheme involving all interactions has been in the centre of discussions in the physics community for many decades. It started more than a century ago as an early vision of Kaluza and Klein \cite{Kaluza:1921, Klein:1926}, who were trying to unify the known interactions at the time, namely gravity and electromagnetism, by going to five dimensions. An interesting revival of the Kaluza-Klein scheme started in the late sixties when it was realised that non-abelian gauge groups, such as those that constitute the well established, today, Standard Model (SM) of Particle Physics, appear in addition to the $U(1)$ of electromagnetism when one considers further extensions of the space dimensions \cite{Kerner:1968, CHO1987358, PhysRevD.12.1711}. Assuming that the total spacetime can be described as a direct product structure, $M_D = M_4\times B$, where $M_4$ is the usual Minkowski spacetime and $B$ is a compact Riemannian space with a non-abelian isometry group $S$, the dimensional reduction of the theory leads to gravity, coupled to a Yang-Mills theory with a gauge group containing $S$ and scalars in four dimensions. Therefore, a very attractive geometrical unification of gravity with other interactions, potentially those of the SM, could be achieved together with a natural explanation of the gauge symmetries. Unfortunately though, serious problems exist in the Kaluza-Klein framework; for instance, there is no classical ground state corresponding to the direct product structure of $M_D$ from the gravity part. On the other hand, from the point of view of Particle Physics comes the most serious problem, that is, after adding fermions to the original action, it is impossible to obtain chiral fermions in four dimensions \cite{Witten:1983}. Eventually these problems are resolved by adding Yang-Mills fields in the original gravity action, at the cost though of abandoning the attractive pure geometric unification of all interactions. Accepting the fact that one has to introduce Yang-Mills fields in higher dimensions, and considering that they are part of a Grand Unified Theory (GUT) containing a Dirac one \cite{Georgi1999, FRITZSCH1975193}, the requirement of obtaining chiral fermions in four dimensions is reduced to the demand that the total dimension of spacetime has to be of the form $4k+2$ (see e.g., ref. \cite{CHAPLINE1982461}).

During the last several decades the Superstring theories (see e.g., refs. \cite{Green2012-ul, polchinski_1998, Lust:1989tj}) have dominated the research on extra dimensions. In particular, heterotic string theory \cite{GROSS1985253}, defined in ten dimensions, was established as the most promising one based on the fact that the SM gauge group and its simple extensions could be accommodated into the GUTs that resulted after the dimensional reduction of the theory, which originally contained an $E_8 \times E_8$ gauge theory. However, it should be noted that  before the formulation of superstring theories, another framework had been developed that focused on the dimensional reduction of higher-dimensional gauge theories, which provided another interesting alternative direction to explore the unification of fundamental interactions \cite{forgacs, MANTON1981502, Kubyshin:1989vd, KAPETANAKIS19924, LUST1985309}. The latter approach to unify fundamental interactions, which shared common objectives with the superstring theories, was first investigated by Forgacs-Manton (F-M) and Scherk-Schwartz (S-S). F-M explored the concept of Coset Space Dimensional Reduction (CSDR) \cite{forgacs, MANTON1981502, Kubyshin:1989vd, KAPETANAKIS19924}, which can naturally lead to chiral fermions, while S-S focused on the group manifold reduction \cite{SCHERK197961}, which, although it does not admit chiral fermions, its basic idea was used later as a prototype in many superstring model building attempts. Recent developments within the CSDR framework and attempts towards realistic models that can be confronted with experiment can be found in the refs. \cite{Manousselis:2000aj, Manousselis:2001xb, Manousselis:2001re, Manousselis_2004, Chatzistavrakidis:2009mh, Irges:2011de, Manolakos:2020cco, Patellis:2024dfl}.

It should be added that, in the attempts to achieve the goal of unification of all interactions, another direction has been developed directly in four dimensions, based on the obvious interface of the notion of gauge invariance. Concretely, the SM of Particle Physics is clearly based on gauge theories, while it is also long known that gravity may, too, be regarded as a gauge theory \cite{utiyama, kibble1961,Sciama, Umezawa, Matsumoto, macdowell, Ivanov:1980tw, Ivanov:1981wm, stellewest, Kibble:1985sn}. Therefore, the further examination of this relationship appeared to be a very interesting challenge. One of the main reasons for the renewed interest in this subject, before the superstrings period, was the progress that was done in supergravity theories (see e.g. \cite{freedman_vanproeyen_2012, Ortín_2015}), which can be regarded very profitably as gauge theories. More recently, the interest was also directed towards noncommutative gravity \cite{castellani, Chatzistavrakidis_2018, Manolakos_paper1, manolakosphd, Manolakos_paper2, Manolakos:2022universe, Manolakos:2023hif, roumelioti2407}.
  
Weyl \cite{weyl, weyl1929} was the first to associate electromagnetism with the phase transformations of the electron field, and the first to develop the vierbein formalism, which appears to be very useful in gauge theories of gravity. A crucial step in the development of gravity as a gauge theory was done by Utiyama \cite{utiyama}, who demonstrated that gravity might be regarded as a gauge theory of the Lorentz group $SO(1,3)$, i.e. the spin connection could be treated as the gauge field of the theory, although the vierbein was introduced in a rather ad hoc way. The latter weakness was improved by Kibble \cite{kibble1961} and Sciama \cite{Sciama} who, instead, considered the gauging of the Poincare group. A further, more elegant development of the theory was done by Stelle and West \cite{stellewest, Kibble:1985sn}, who identified the spin connection and the veilbein as parts of the gauge fields of the de Sitter (dS) $SO(1,4)$, or the Anti-de Sitter (AdS) group $SO(2,3)$, which were spontaneously broken by a scalar field to the Lorentz $SO(1,3)$ group. It is also worth noting that the gauge theory of the conformal group $SO(2,4)$ was used in constructing Weyl Gravity (WG) \cite{KAKU1977304, Roumelioti:2024lvn, Fradkin:1985am}, Fuzzy Gravity (FG) \cite{Chatzistavrakidis_2018, Manolakos_paper1, manolakosphd, Manolakos_paper2, Manolakos:2022universe, Manolakos:2023hif, roumelioti2407}, and its supersymmetric extension, the superconformal group, in $N = 1$ supergravity \cite{freedman_vanproeyen_2012, KAKU1977304}. Extensive studies on WG have also been done from other points of view (see e.g. \cite{Fradkin:1985am, Maldacena:2011mk, mannheim, Anastasiou:2016jix, ghilencea2023, Hell:2023rbf, Condeescu:2023izl}).

Another interesting and more direct suggestion, towards unifying gravity as gauge theory with the other known interactions described by GUTs, has been suggested in the past \cite{Percacci:1984ai, Percacci_1991} and revived recently \cite{Weinberg:1984ke,Nesti_2008, Nesti_2010, Krasnov:2017epi, Chamseddine2010, Chamseddine2016, noncomtomos, Konitopoulos:2023wst}. It is based on the observation that the tangent group of a curved manifold does not necessarily have the same dimension as the manifold \cite{Weinberg:1984ke}. This possibility opens the very interesting avenue that one can consider higher than four dimensional tangent groups in a four-dimensional spacetime and possibly achieve unification of gravity with internal interactions by gauging these higher-dimensional tangent groups. Then, to a great extent, the machinery that has been developed to examine higher-dimensional theories with extra physical space dimensions, such as those used in the CSDR scheme \cite{CHAPLINE1982461, forgacs, MANTON1981502, Kubyshin:1989vd, KAPETANAKIS19924, LUST1985309, SCHERK197961, Manousselis_2004, Chatzistavrakidis:2009mh, Irges:2011de, Manolakos:2020cco, Patellis:2024dfl}, can be transferred to the four-dimensional constructions, since they have the same tangent group. Examples of the latter are the constraints, such as the Weyl condition, that have to be imposed in the higher-dimensional theories in order to result in realistic four-dimensional chiral theories describing the internal interactions after the dimensional reduction \cite{CHAPLINE1982461, KAPETANAKIS19924}. Similarly, the imposition of the Majorana condition, in addition to the Weyl, is needed in the extra-dimensional theories, in order to avoid a possible doubling of the spectrum of the reduced chiral theories \cite{CHAPLINE1982461, KAPETANAKIS19924}. Along this outlined direction, a unification of the gauge conformal group with those of the internal interactions has recently been constructed \cite{Konitopoulos:2023wst, Roumelioti:2024lvn}. Subsequently, this construction was extended to the unification of the gauge conformal group on a covariant, noncommutative (fuzzy) space with internal interactions \cite{roumelioti2407}.

Consequently, we shall also present our gauge-theoretic construction of the four-dimensional matrix model of noncommutative gravity that has been done so far. In order to build a noncommutative gauge theory, a background noncommutative space is required to accommodate it (for a review, see e.g., ref. \cite{Manolakos:2022universe}). Therefore, first we focus on the construction of a four-dimensional covariant noncommutative space, which plays the role of the background space, and then we present the gravity model, constructed as a noncommutative gauge theory on the above space.

It should be emphasised that formulating gravity in the noncommutative framework is a quite difficult task, given that and common noncommutative deformations break Lorentz invariance. It is very interesting though to note that there are known certain types of deformations constituting a class of covariant noncommutative spacetimes, which preserve the spacetime isometries. The basic characteristic of this class of noncommutative spaces is that the generators of the used Lie algebra, in principle related to the isometries of the spacetimes, are identified with the coordinates. In this way, it is natural to obtain coordinate noncommutativity as well as covariance. Snyder \cite{Snyder:1946qz} was the first to associate the position operators with elements of a Lie algebra already in 1947 and constructed this way a Lorentz invariant and discrete spacetime due to the introduced unit length in his construction. Snyder’s position operators belong to the four-dimensional de Sitter algebra, $SO(1,4)$. Then Yang \cite{yang1947} enlarged the $SO(1,4)$ to $SO(1,5)$ aiming to allow for continuous translations in Snyder’s framework, while finally Kastrup \cite{kastrup_1966} further enlarged the starting Lie algebra to the four-dimensional conformal one, $SO(2,4)$. In the recent years the above early works were reconsidered in the framework of noncommutative geometry, with possible applications in string theory, by Heckman-Verlinde \cite{Heckman_2015}, where the authors build a general conformal field theory defined on covariant noncommutative versions of the four-dimensional de Sitter (dS) or anti-de Sitter (AdS) spacetime. Therefore, covariant noncommutative spacetimes appear to overcome the problem of breaking Lorentz invariance, with which the constructions of the most common noncommutative spaces were usually confronted. They preserve Lorentz invariance, providing moreover a short-distance cutoff. In our constructions, we took a natural step further, promoting the Lorentz invariance of the covariant noncommutative spacetime to an associated local symmetry, which can describe the corresponding gravity theory on such a space \cite{Chatzistavrakidis_2018,Manolakos_paper1,manolakosphd,Manolakos_paper2}.  

A very interesting and relatively simple noncommutative space is the Fuzzy Sphere, $S^2_F$, which is the discrete matrix approximation of the ordinary, continuous, sphere, $S^2$, with the property of preserving its isometries \cite{Madore_1992}, i.e. the Fuzzy Sphere is also covariant. Unfortunately, the Fuzzy Sphere cannot be extended in higher dimensions in a straightforward way. Nevertheless, the suggestion resulting from the past experience described above, is that in order to formulate a fuzzy covariant noncommutative space, one has to employ a wider symmetry, containing all generators of the isometry of the continuous one, identified as coordinates of the noncommutative space under construction. Then in the case of the fuzzy four-dimensional de Sitter spacetime, $dS_4$, that has been constructed, instead of using the algebra of its isometry group generators, $SO(1,4)$, we have considered a larger one in which the latter could be embedded. The simplest enlarged algebras of $SO(1,4)$ are the $SO(1,5)$ and the conformal, $SO(2,4)$. For instance, the algebra $SO(1,5)$ contains all the generators of the subalgebra $SO(1,4)$ and the commutation relations of the latter are covariant as seen in the larger group. This is the way, along which we have successfully constructed a four-dimensional FG theory \cite{Manolakos:2022universe, Manolakos_paper1, Manolakos_paper2, Manolakos:2023hif}.

In the present article, we review the basic features of Conformal Gravity (CG) and FG as well as their Unification with internal interactions based on $SO(10)$ GUT \cite{Roumelioti:2024lvn,roumelioti2407} (see also \cite{Patellis:2024znm}).

\section{Conformal Gravity}

As was discussed in Section \ref{intro}, Einstein Gravity (EG) has been treated as the gauge theory of the Poincar\'{e} group, though much insight and elegance was gained by considering instead the gauging of the dS, $SO(1,4)$, and the AdS, $SO(2,3)$. Both of these groups contain the same number of generators as the Poincar\'{e} group, i.e. 10, and can be spontaneously broken by a scalar field to the Lorentz group, $SO(1,3)$ \cite{stellewest, Kibble:1985sn, Roumelioti:2024lvn, manolakosphd}. The Poincar\'{e}, the dS and the AdS groups are all subgroups of the conformal group $SO(2,4)$, which has 15 generators and is the group of transformations on space-time which leave invariant the null interval $ds^2 = \eta_{\mu\nu} dx^\mu dx^\nu = 0$. In ref \cite{Kaku:1978nz} the gauge theory formalism of Gravity was extended to the conformal group constructing in this way the CG. The breaking of CG to EG or to Weyl’s scale invariant theory of gravity was done via the imposition of constraints (see e.g. \cite{Kaku:1978nz}). However, in ref \cite{Roumelioti:2024lvn}, for the first time, the breaking of the conformal gauge group was done spontaneously, induced by the introduction of a scalar field in the action, which obtains non-vanishing vev, using the Lagrange multiplier method. 

\subsection{Spontaneous symmetry breaking}
The gauge group, $SO(2,4)$, consists of fifteen generators. In four-dimensional notation they correspond to six Lorentz transformations, $M_{ab}$, four translations, $P_a$, four special conformal transformations (conformal boosts), $K_a$, as well as the dilatation, $D$.

The gauge connection, $A_\mu$, as an element of the $SO(2,4)$ algebra, can be expanded in terms of the generators as
\begin{equation}
A_\mu= \frac{1}{2}\omega_\mu{}^{a b} M_{a b}+e_\mu{}^a P_a+b_\mu{}^a K_a+\tilde{a}_\mu D,
\end{equation}
where, for each generator a gauge field has been introduced. The gauge field related to the translations is identified as the vierbein, while the one of the Lorentz transformations is identified as the spin connection. The field strength tensor is of the form
\begin{equation}\label{fst}
F_{\mu \nu}=\frac{1}{2}R_{\mu \nu}{}^{a b} M_{a b}+\tilde{R}_{\mu \nu}{}^a P_a+R_{\mu \nu}{}^a K_a+R_{\mu \nu} D,
\end{equation}
where
\begin{equation}\label{curves}
\begin{aligned}
R_{\mu \nu}{}^{a b} & =\partial_\mu \omega_\nu{}^{a b}-\partial_\nu \omega_\mu{}^{a b}-\omega_\mu{}^{a c} \omega_{\nu c}{}^b+\omega_\nu{}^{a c} \omega_{\mu c}{}^b-8 e_{[\mu}{}^{[a} b_{\nu]}{}^{b]} \\
& =R_{\mu \nu}^{(0) a b}-8 e_{[\mu}{}^{[a} b_{\nu]}{}^{b]}, \\
\tilde{R}_{\mu \nu}{}^a & =\partial_\mu e_\nu{}^a-\partial_\nu e_\mu{}^a+\omega_\mu{}^{a b} e_{\nu b}-\omega_\nu{}^{a b} e_{\mu b}-2 \tilde{a}_{[\mu} e_{\nu]}{}^a \\
& =T_{\mu \nu}^{(0) a}(e)-2 \tilde{a}_{[\mu} e_{\nu]}{}^a, \\
R_{\mu \nu}{}^a & =\partial_\mu b_\nu{}^a-\partial_\nu b_\mu{}^a+\omega_\mu{}^{a b} b_{\nu b}-\omega_\nu{}^{a b} b_{\mu b}+2 \tilde{a}_{[\mu} b_{\nu]}{}^a\\
&=T_{\mu \nu}^{(0) a}(b)+2 \tilde{a}_{[\mu} b_{\nu]}{}^a,\\
R_{\mu \nu} & =\partial_\mu \tilde{a}_\nu-\partial_\nu \tilde{a}_\mu+4 e_{[\mu}{}^a b_{\nu] a},
\end{aligned}
\end{equation}
where $T_{\mu \nu}^{(0) a}(e)$ and $R_{\mu \nu}^{(0) a b}$ are the torsion and curvature component tensors in the four-dimensional vierbein formalism of General Relativity (GR), while $T_{\mu \nu}^{(0) a}(b)$ is the torsion tensor related to the gauge field $b_\mu{}^a$. 

We shall start by choosing the parity conserving action, which is quadratic in terms of the field strength tensor \eqref{fst}, in which we have introduced a scalar that belongs to the adjoint rep, $15$, of $SO(6) \sim SO(2,4)$ along with a dimensionful parameter, $m$:
\begin{equation}
    S_{SO(2,4)}=a_{CG}\int d^4x [\operatorname{tr} \epsilon^{\mu \nu \rho \sigma} m\phi F_{\mu \nu}F_{\rho \sigma}+(\phi^2-m^{-2} \mathbb{1}_4)], 
\end{equation}
where the trace is defined as $ \operatorname{tr}\rightarrow \epsilon_{abcd} [\text{Generators}]^{abcd}$.

The scalar expanded on the generators is:
\begin{equation}
\phi=\phi^{a b} M_{a b}+\tilde{\phi}^a P_a+\phi^a K_a+\tilde{\phi} D,
\end{equation}

In accordance with \cite{Li:1973mq}, we pick the specific gauge in which $\phi$ is diagonal of the form $\operatorname{diag}(1,1,-1,-1)$. Specifically we choose $\phi$ to be only in the direction of the dilatation generator $D$:
\begin{equation}
    \phi=\phi^0=\tilde{\phi}D \xrightarrow{\phi^2=m^{-2}\mathbb{1}_4}\phi=-2m^{-1} D.
\end{equation}
In this particular gauge the action reduces to
\begin{equation}
    S=-2a_{CG}\int d^4x \operatorname{tr} \epsilon^{\mu \nu \rho \sigma} F_{\mu \nu}F_{\rho \sigma}D, 
\end{equation}
and the gauge fields $e,b$ and $\tilde{a}$ become scaled as $me,mb$ and $m\tilde{a}$ correspondingly. After straightforward calculations, using the expansion of the field strength tensor as in eq. \eqref{fst}, and the anticommutation relations of the generators, we obtain: 
\begin{equation}
\begin{gathered}
    S=-2a_{CG}\int d^4x \operatorname{tr} \epsilon^{\mu \nu \rho \sigma}\Big[\frac{1}{4}R_{\mu \nu}{}^{ab}R_{\rho \sigma}{}^{cd}M_{ab}M_{cd}D+\\
    +i\epsilon_{abcd}(R_{\mu \nu}{}^{ab}R_{\rho \sigma}{}^{c} K^d D - R_{\mu \nu}{}^{ab}\tilde{R}_{\rho \sigma}{}^{c}P^{d}D)+\\+(\frac{1}{2}\tilde{R}_{\mu \nu}{}^{a}R_{\rho \sigma} + 2\tilde{R}_{\mu \nu}{}^{a}R_{\rho \sigma}{}^{b})M_{ab}+\\
    +(\frac{1}{4}R_{\mu \nu}R_{\rho \sigma}- 2\tilde{R}_{\mu \nu}{}^{a}R_{\rho \sigma a})D   
    \Big].
\end{gathered}
\end{equation}
In this point we employ the trace on the several generators and their products. In particular:
\begin{equation}
\begin{gathered}
   \operatorname{tr}[K^{d}D]=\operatorname{tr}[P^{d}D]=\operatorname{tr}[M_{ab}]= \operatorname{tr}[D]=0, \\
        \text{and}\quad \operatorname{tr}[M_{ab}M_{cd}D]=-\frac{1}{2}\epsilon_{abcd}.
\end{gathered}
\end{equation}
The resulting broken action is:
\begin{equation}
\label{BrokenActionConformal}
     S_{\mathrm{SO}(1,3)}=\frac{a_{CG}}{4}\int d^4x \epsilon^{\mu \nu \rho \sigma}\epsilon_{abcd}R_{\mu \nu}{}^{ab}R_{\rho \sigma}{}^{cd},
\end{equation}
while its invariance has obviously been reduced only to Lorentz. Before continuing, we notice that there is no term containing the field $\tilde{a}_\mu$ present in the action. Thus, we may set $\tilde{a}_\mu=0$. This simplifies the form of the two component field strength tensors related to the $P$ and $K$ generators:
\begin{equation}
\begin{aligned}
  \tilde{R}_{\mu \nu}{}^a & =mT_{\mu \nu}^{(0) a}(e)-2 m^2\tilde{a}_{[\mu} e_{\nu]}{}^a \longrightarrow mT_{\mu \nu}^{(0) a}(e), \\
R_{\mu \nu}{}^a &=mT_{\mu \nu}^{(0) a}(b)+2m^2 \tilde{a}_{[\mu} b_{\nu]}{}^a \longrightarrow mT_{\mu \nu}^{(0) a}(b).
\end{aligned}
\end{equation}
The absence of the above field strength tensors in the action, allows us to also set $\tilde{R}_{\mu \nu}{}^a=R_{\mu \nu}{}^a=0$, and thus to obtain a torsion-free theory. Since $R_{\mu \nu}$ is also absent from the expression of the broken action, it may also be set equal to zero. From its definition in eq. \eqref{curves}, then we obtain the following relation among $e$ and $b$: 
\begin{equation}\label{ef}
    e_\mu{}^a b_{\nu a}-e_{\nu}{}^{a}b_{\mu a}=0.
\end{equation}
The above result reinforces one to consider solutions that relate $e$ and $b$.
We examine two possible solutions of eq. \eqref{ef}.

\subsubsection*{When $b_\mu{}^a = ae_\mu{}^a$ - Einstein-Hilbert (E-H) action in the presence of a cosmological constant} In this case, first proposed in \cite{Chamseddine:2002fd}, by a simple substitution we obtain:
\begin{equation}
\begin{aligned}
        S_{\mathrm{SO}(1,3)} =\frac{a_{CG}}{4}\int d^4 x \epsilon^{\mu \nu \rho \sigma} \epsilon_{a b c d}&\left[R_{\mu \nu}^{(0) a b}-4m^2a\left(e_\mu{}^a e_\nu{}^b-e_\mu{}^b e_\nu{}^a\right)\right]\\
       &\left[R_{\rho \sigma}^{(0) c d}-4m^2a\left(e_\rho{}^c e_\sigma{}^d-e_\rho{}^d e_\sigma{}^c\right)\right]\longrightarrow\\\\
        S_{\mathrm{SO}(1,3)} =\frac{a_{CG}}{4}\int d^4 x \epsilon^{\mu \nu \rho \sigma} \epsilon_{a b c d}&[R_{\mu \nu}^{(0) a b}-8m^2ae_\mu{}^a e_\nu{}^b]\\
        &\qquad\qquad\left[R_{\rho \sigma}^{(0) c d}-8m^2ae_\rho{}^c e_\sigma{}^d\right],
       \end{aligned}
\end{equation}
which yields
\begin{equation}\label{so24finalaction}
\begin{aligned}
    S_{\mathrm{SO}(1,3)}=\frac{a_{CG}}{4}\int d^4 x \epsilon^{\mu \nu \rho \sigma} \epsilon_{a b c d}[R_{\mu \nu}^{(0) a b}R_{\rho \sigma}^{(0) c d}-16m^2aR_{\mu \nu}^{(0) a b}e_\rho{}^c e_\sigma{}^d+\\
    +64m^4a^2 e_\mu{}^a e_\nu{}^b e_\rho{}^c e_\sigma{}^d].
\end{aligned}
\end{equation}

This action consists of three terms: a Gauss-Bonnet topological term, the E-H action, and a cosmological constant term. For $a<0$, the above describes GR in AdS space.

\subsubsection*{When $b_\mu{}^a = -\frac{1}{4} (R_\mu{}^a - \frac{1}{6} R e_\mu{}^a)$ - Weyl action}
This relation between $b$ and $e$, which is again a solution of \eqref{ef}, was suggested in refs \cite{Kaku:1978nz} and \cite{freedman_vanproeyen_2012}. Taking this into account we obtain the following action:
\begin{equation}
    \begin{aligned}
             S_W=\frac{a_{CG}}{4}&\int d^4 x \epsilon^{\mu \nu \rho \sigma} \epsilon_{a b c d}
             \\&\left[R_{\mu \nu}^{(0) a b}+\frac{1}{2}\left(m e_\mu{}^{[a} R_\nu{}^{b]}-me_\nu{}^{[a} R_\mu{}^{b]}\right)-\frac{1}{3} m^2 R e_\mu{}^{[a} e_\nu{}^{b]}\right]\\
             &\left[R_{\rho \sigma}^{(0) c d}+\frac{1}{2}\left(m e_\rho{}^{[c} R_\sigma{}^{d]}-m e_\sigma{}^{[c} R_\rho{}^{d]}\right)-\frac{1}{3} m^2 R e_\rho{}^{[c} e_\sigma{}^{d]}\right].
    \end{aligned}
\end{equation}
Considering the rescaled vierbein $\tilde{e}_\mu{}^{a}=m e_\mu{}^{a}$ and recalling that $R_{\mu \nu}^{(0) a b}=-R_{\nu \mu}^{(0) a b}$, we obtain
\begin{equation}
    \begin{aligned}
             S_W=\frac{a_{CG}}{4}\int d^4 x&\epsilon^{\mu \nu \rho \sigma} \epsilon_{a b c d}\\&\left[R_{\mu \nu}^{(0) a b}-\frac{1}{2}\left(\tilde{e}_\mu{}^{[a} R_\nu{}^{b]}-\tilde{e}_\nu{}^{[a} R_\mu{}^{b]}\right)+\frac{1}{3} R \tilde{e}_\mu{}^{[a} \tilde{e}_\nu{}^{b]}\right]\\
             &\left[R_{\rho \sigma}^{(0) c d}-\frac{1}{2}\left( \tilde{e}_\rho{}^{[c} R_\sigma{}^{d]}- \tilde{e}_\sigma{}^{[c} R_\rho{}^{d]}\right)+\frac{1}{3} R \tilde{e}_\rho{}^{[c} \tilde{e}_\sigma{}^{d]}\right],
    \end{aligned}
\end{equation}
which is equal to 
\begin{align}
             S_W=\frac{a_{CG}}{4}\int d^4 x \epsilon^{\mu \nu \rho \sigma} \epsilon_{a b c d}C_{\mu \nu}{}^{a b}C_{\rho \sigma}{}^{c d},
\end{align}
where $C_{\mu \nu}{}^{a b}$ is the Weyl conformal tensor. This action leads to the well-know four-dimensional scale invariant Weyl action,
\begin{equation}
S_W =2a_{CG}\int \mathrm{d}^4 x\left(R_{\mu \nu} R^{\nu \mu}-\frac{1}{3} R^2\right).
\end{equation}

\section{Noncommutative Gauge Theory of 4D Gravity - Fuzzy Gravity}
\subsection{The Background Space}

Before we move on with the gauge theory of FG, we will first have to establish the background space, on which this theory will be formulated. In refs \cite{yang1947, Heckman_2015,Manolakos_paper1, Manolakos_paper2, Manolakos:2022universe} extending the original Snyder's suggestion \cite{Snyder:1946qz} the authors have considered the group the $SO(1,5)$ and have assigned the 4-d spacetime coordinates to elements of its Lie algebra.

More specifically starting with the group $SO(1,5)$, whose generators obey the following Lie algebra:
\begin{equation}
     \left[J_{mn},J_{rs}\right]=i\left(\eta_{mr}J_{ns}+\eta_{ns}J_{mr}-\eta_{nr}J_{ms}-\eta_{ms}J_{nr}\right),
\end{equation}
where $m,n,r,r = 0,\ldots, 5$, and $\eta_{mn} = diag(-1, 1, 1, 1, 1, 1)$. Performing the decompositions of $SO(1, 5)$ to its maximal subgroups, up to $SO(1, 3)$, i.e., $SO(1,5) \supset  SO(1, 4)$ and $SO(1, 4) \supset SO(1, 3)$, turns the above commutation relation to the following:
\begin{equation}
\begin{gathered}  
    \left[J_{ij},J_{kl}\right]=i\left(\eta_{i k}J_{j l}+\eta_{j l}J_{i k}-\eta_{j k}J_{i l}-\eta_{i l}J_{j k}\right),\\
    \left[J_{i j},J_{k5}\right]=i\left(\eta_{i k}J_{j5}-\eta_{j k}J_{i5}\right),\\
    \left[J_{i5},J_{j5}\right]=i J_{ij},\\
    \left[J_{i j},J_{k4}\right]=i\left(\eta_{i k}J_{j4}-\eta_{j k}J_{i4}\right),\\
    \left[J_{i4},J_{j4}\right]=i J_{ij},\\
    \left[J_{i4},J_{j5}\right]=i \eta_{ij}J_{45},\\
    \left[J_{i j},J_{45}\right]=0,\\
    \left[J_{i 4},J_{45}\right]=-i J_{i5},\\
    \left[J_{i 5},J_{45}\right]=i J_{i4}.
\end{gathered}
\end{equation}
Next one may convert the generators to physical quantities by setting
  \begin{equation}
    \Theta_{ij}=\hbar J_{ij}, \, \text{and} \ X_i=\lambda J_{i5},
\end{equation}
where $\lambda$ is a natural unit of length, and furthermore identify the momenta as
\begin{equation}
\label{identifications.P}
    P_i=\frac{\hbar}{\lambda}J_{i4},
\end{equation}
and set $h=J_{45}$. Then given these identifications and the commutation relations above, one obtains:
\begin{equation}
\begin{gathered}  
    \left[\Theta_{ij},\Theta_{kl}\right]=i \hbar\left(\eta_{i k}\Theta_{j l}+\eta_{j l}\Theta_{i k}-\eta_{j k}\Theta_{i l}-\eta_{i l}\Theta_{j k}\right),\\
    \left[\Theta_{i j},X_{k}\right]=i\hbar \left(\eta_{i k}X_{j}-\eta_{j k}X_{i}\right),\\
    \left[\Theta_{i j},P_{k}\right]=i\hbar\left(\eta_{i k}P_{j}-\eta_{j k}P_{i}\right),\\
    \left[X_{i},X_{j}\right]=\frac{i\lambda^2}{\hbar} \Theta_{ij},\ 
    \left[X_{i},P_{j}\right]=i\hbar \eta_{ij}h,\ 
    \left[P_{i},P_{j}\right]=\frac{i \hbar}{\lambda^2} \Theta_{ij},\\
    \left[X_{i},h\right]=\frac{i\lambda^2}{\hbar} P_{i},\ 
    \left[P_{i},h\right]=-\frac{i\hbar}{\lambda^2} X_{i},\ 
    \left[\Theta_{i j},h\right]=0.
\end{gathered}
\end{equation}
From the above relations, it becomes clear that one is led to the following significant results. First since the coordinates as well as the momenta are elements of this Lie algebra, they exhibit a noncommutative behavior, implying that both the space-time and the momentum space become quantized. In addition it becomes evident that the commutation relation between coordinates and momenta naturally yields a Heisenberg-type uncertainty relation.

\subsection{Gauge Group and Representation}
Starting with the formulation of a gauge theory for gravity in the space mentioned above, we first have to determine the group that will be gauged. Naturally, the group we will choose is the one that describes the symmetries of the theory, in this case, the isometry group of $dS_4$, $SO(1,4)$. Since, as shown in refs. \cite{Manolakos_paper1, Manolakos_paper2}, in noncommutative gauge theories the use of anticommutators of the algebra generators is inevitable, and since the anticommutators of the generators of either of the above isometry group do not necessarily yield elements that belong in the same algebra, we have to take into account the closing of the anticommutators of the relevant gauge group generators. In order to achieve that, we have to pick a specific representation of the algebra generators, and subsequently extend the initial gauge group to one with larger symmetry, in which both the commutator and anticommutator algebras close. Following this procedure, we are led to the extension of our initial gauge group $SO(1,4)$ to $SO(2,4) \times U(1)$.

\subsection{Fuzzy Gravity}

Since we have already determined the appropriate gauge group of the theory, we are now able to move on with the gauging procedure on the fuzzy space that was presented above. Following the steps described in \cite{Manolakos_paper1}, we firstly have to introduce the covariant coordinate of the theory, which is defined as:
\begin{equation}\label{CovariantCoordinate}
    \mathcal{X}_\mu=X_\mu \otimes \mathbb{1}_4 +A_\mu (X)\, ,
\end{equation}
where $A_\mu$ is the gauge connection of the theory. The gauge connection can, in turn, be expanded on the gauge group generators as:
\begin{equation}\label{GaugeConnectionFuzzy}
    A_\mu = a_\mu \otimes \mathbb{1}_{4} +  \omega_\mu{}^{ab}\otimes M_{ab}+ e_\mu{}^{a}\otimes P_a + b_\mu{}^{a}\otimes K_a + \tilde{a}_\mu\otimes D\,.
\end{equation}
Given the above expansion \eqref{GaugeConnectionFuzzy}, the explicit form of the covariant coordinate \eqref{CovariantCoordinate} can be written down as:
\begin{equation}
    \mathcal{X}_\mu= (X_\mu + a_\mu) \otimes \mathbb{1}_{4} +  \omega_\mu{}^{ab}\otimes M_{ab}+ e_\mu{}^{a}\otimes P_a + b_\mu{}^{a}\otimes K_a + \tilde{a}_\mu\otimes D\,.
\end{equation}
At this point, what remains to be determined is the appropriate covariant field strength tensor for the theory. In the case of noncommutativity, the latter is defined as \cite{Madore_1992, Manolakos_paper1}:
\begin{equation}
    \hat{F}_{\mu \nu}\equiv\left[\mathcal{X}_{\mu}, \mathcal{X}_{\nu}\right]-\kappa^2 \hat{\Theta}_{\mu \nu}\, ,
\end{equation}
where $\hat{\Theta}_{{\mu}{\nu}}\equiv\Theta_{{\mu}{\nu}}+\mathcal{B}_{{\mu}{\nu}}$, in which $\mathcal{B}_{{\mu}{\nu}}$ is a 2-form field taking care of the transformation of $\Theta$, promoting it to its covariant form. Since $\hat{F}_{\mu \nu}$ is an element of the gauge algebra it can also be expanded on the algebra's generators as 
\begin{equation}
    \hat{F}_{\mu \nu}= R_{\mu \nu} \otimes \mathbb{1}_4 +\frac{1}{2} R_{\mu \nu}{}^{a b} \otimes M_{a b} + \tilde{R}_{\mu \nu}{}^{a} \otimes P_{a}+R_{\mu \nu}{}^{a} \otimes K_{a}+\tilde{R}_{\mu \nu} \otimes D\,.
\end{equation}
The SSB goes along the same lines as the one described in the case of Conformal Gravity, i.e. we introduce a scalar field, $\Phi(X)$, belonging the 2nd rank antisymmetric rep of $SO(2,4)$, in the action and fix it in the gauge that leads to the Lorentz group (see \cite{Manolakos_paper1, Manolakos_paper2, Roumelioti:2024lvn}). This scalar field must be charged under the $U(1)$ gauge symmetry so that the $U(1)$ part breaks and it doesn't appears in the final symmetry. Introducing the scalar field, the action takes the form:
\begin{equation}
\mathcal{S}=\operatorname{Trtr} \Big[\lambda \Phi(X) \varepsilon^{\mu \nu \rho \sigma}\hat{F}_{\mu \nu} \hat{F}_{\rho \sigma} +\eta\left(\Phi(X)^2-\lambda^{-2} \mathbb{1}_N \otimes \mathbb{1}_4\right)\Big],
\end{equation}
where the first trace is over the coordinate matrices, the second is over the generators of the gauge group, $\eta$ is a Lagrange multiplier, and $\lambda$ is a dimensionfull parameter. The scalar field itself is also an element of the gauge group, and hence can be expanded on its generators, as
\begin{equation}
\begin{aligned}
\Phi(X)=\phi(X) &\otimes \mathbb{1}_4+\phi^{a b}(X) \otimes M_{a b}+\tilde{\phi}^a(X) \otimes P_a+\\&+\phi^a(X) \otimes K_a+\tilde{\phi}(X) \otimes D.
\end{aligned}
\end{equation}
As mentioned above, just like in our previous works  \cite{Manolakos_paper1, Manolakos_paper2, Roumelioti:2024lvn}, we gauge-fix the scalar field into the dilatation direction:
\begin{equation}
\Phi(X)=\left.\tilde{\phi}(X) \otimes D\right|_{\tilde{\phi}=-2 \lambda^{-1}}=-2 \lambda^{-1} \mathbb{1}_N \otimes D.
\end{equation}
On-shell, when the above equation holds, the aforementioned action reduces to the following form:
\begin{equation}
\mathcal{S}_{b r}=\operatorname{Tr}\left(\frac{\sqrt{2}}{4} \varepsilon_{a b c d} R_{m n}{}^{a b} R_{r s}{}^{c d}-4 R_{m n} \tilde{R}_{r s}\right) \varepsilon^{m n r s},
\end{equation}
while any other term, along with the Lagrange multiplier, has vanished due to the gauge fixing. This resulting action now bears the remaining $SO(1,3)$ gauge symmetry, following the SSB. Moreover, when the commutative limit of the above action is considered (for details, see \cite{Manolakos_paper2}), it reduces to the Palatini action, which in turn is equivalent to EG, with a cosmological constant term present.

\section{Unification of Conformal and Fuzzy Gravities with Internal Interactions, Fermions and Breakings}

In \cite{Roumelioti:2024lvn}, it was suggested that the unification of the CG with internal interactions based on a framework that results in the GUT $SO(10)$ could be achieved using the $SO(2, 16)$ as unifying gauge group. As it was emphasized in Section \ref{intro} the whole strategy was based on the observation that the dimension of the tangent space is not necessarily equal to the dimension of the corresponding curved manifold \cite{roumelioti2407, Percacci:1984ai, Percacci_1991, Nesti_2008, Nesti_2010, Krasnov:2017epi, Chamseddine2010, Chamseddine2016, noncomtomos, Konitopoulos:2023wst, Weinberg:1984ke}. An additional fundamental observation \cite{Roumelioti:2024lvn} is that in the case of $SO(2, 16)$ one can impose Weyl and Majorana conditions on fermions \cite{D_Auria_2001, majoranaspinors}. More specifically, using Euclidean signature for simplicity (the implications of using non-compact space are explicitly discussed in \cite{Roumelioti:2024lvn}), one starts with $SO(18)$ and with the fermions in its spinor representation, $256$. Then the SSB of $SO(18)$ leads to its maximal subgroup $SO(6)\times SO(12)$. Let us recall for convenience the branching rules of the relevant reps \cite{Slansky:1981yr, Feger_2020, Li:1973mq},
\begin{equation}\label{so18}
\begin{aligned}
S O(18) & \supset S O(6) \times S O(12) & & \\
{2 5 6} & =({4}, \overline{{32}})+(\overline{{4}}, {32}) & & \text { spinor } \\
{1 7 0} & =({1}, {1})+({6}, {1 2})+\left({2 0}^{\prime}, {1}\right)+({1}, {7 7}) & & \text { 2nd rank symmetric }
\end{aligned}  
\end{equation}
The breaking of $SO(18)$ to $SO(6) \times SO(12)$ is done by giving a vev to the $\langle 1, 1\rangle$ component of a scalar in the ${170}$ rep. In turn, given that the Majorana condition can be imposed, due to the non-compactness of the used $SO(2, 16) \sim SO(18)$, we are led after the SSB to the $SO(6) \times SO(12)$ gauge theory with fermions in the $(4, \overline{32})$ representation. Then, according to \cite{Roumelioti:2024lvn}, the following SSBs can be achieved by using scalars in the appropriate representations
\begin{equation}
\begin{aligned}
&S O(6) \rightarrow S U(2) \times S U(2),
\end{aligned}
\end{equation}
in the CG sector, and
\begin{equation}
\begin{aligned}
&S O(12) \rightarrow S O(10) \times[U(1)]_{g l o b a l}
\end{aligned}
\end{equation}
in the internal gauge symmetry sector, with fermions in the $16_L(-1)$ under the $SO(10) \times [U(1)]_{global}$. The other generations are introduced as usual with more chiral fermions in the 256 rep of $SO(18)$. We choose scalars in the 2nd rank antisymmetric 15 rep of $SO(6)$ to break the CG gauge group, while the internal interactions gauge group $SO(12)$ is broken spontaneously by scalars in the 77 rep. The 15 rep can be drawn from the $SO(18)$ rep 153:
\begin{equation}\label{153}
153=(15,1)+(6,12)+(1,66),
\end{equation}
while from \ref{so18} we see that the 77 rep can result from a 170 rep of the parent group. Thus, in $SO(6) \times SO(12)$ notation, the scalars breaking the two gauge groups belong to $(15, 1)$ and $(1, 77)$, respectively. According to the above picture we start from some high scale where the $SO(18)$ gauge group breaks, eventually obtaining EG and $SO(10) \times [U(1)]_{\text{global}}$ after several symmetry breakings. From that point, we use the symmetry breaking paths and field content followed in [67], in order to finally arrive at the SM. In particular, the $SO(10)$ group breaks spontaneously into an intermediate group which eventually breaks into the SM gauge group. The intermediate groups are the Pati-Salam (PS) gauge group, $SU(4)_C \times SU(2)_L \times SU(2)_R$, with or without a discrete left-right symmetry, $D$, and the minimal left-right gauge group (LR), $SU(3)_C \times SU(2)_L \times SU(2)_R \times U(1)_{B-L}$, again with or without the discrete left-right symmetry. We will denote the four intermediate gauge groups as $422$, $422D$, $3221$ and $3221D$, respectively. The $SO(10)$ group breaks with a scalar 210 into the $422$ and the $3221D$ groups, with a scalar 54 into $422$ and with a scalar 45 into $3221$. The spontaneous breaking into the SM gauge group from each and every intermediate group is achieved with scalars that are accommodated in a 126 rep (still in $SO(10)$ language), while the Higgs boson necessary for the electroweak breaking will be accommodated in a 10 rep. We will call the scale at which the $SO(10)$ gauge group breaks GUT scale, $M_{GUT}$ , in the sense that all three gauge couplings are unified at that scale, while we will call the scale at which the $422(D)/3221(D)$ groups break intermediate scale, $M_I$. Thus, the consecutive breakings in each case can be seen in Table \ref{table:1}.
\begin{table}\scriptsize
\begin{center}
\begin{tabular}{|c|}
 \hline
 $422:\left.\mathrm{SO}(10)\right|_{M_\text{GUT}} \xrightarrow{\left\langle {2 1 0}_{ {H}}\right\rangle}S U(4)_C \times S U(2)_R \times\left. S U(2)_R\right|_{M_I} \xrightarrow{\left\langle\overline{{1 2 6}}_{ {H}}\right\rangle} \mathrm{SM}$\\ [1.5ex]
 \hline
$422 \mathrm{D}: \left.\mathrm{SO}(10)\right|_{M_\text{GUT}} \xrightarrow{\left\langle {5 4}_{ {H}}\right\rangle} SU(4)_C \times SU(2)_R \times S U(2)_R \times\left.\mathcal{D}\right|_{M_I} \xrightarrow{\left\langle\overline{{1 2 6}}_{ {H}}\right\rangle} \mathrm{SM}$\\[1.5ex]
  \hline
$3221: \left.\mathrm{SO}(10)\right|_{M_\text{GUT}} \xrightarrow{\left\langle {4 5}_{ {H}}\right\rangle}SU(3)_C \times S U(2)_L \times S U(2)_R \times\left. U(1)_{B-L}\right|_{M_I} \xrightarrow{\left\langle\overline{126}_{ {H}}\right\rangle} \mathrm{SM}$\\ [1.5ex]
  \hline
 $3221 \mathrm{D}: \left.\mathrm{SO}(10)\right|_{M_\text{GUT}} \xrightarrow{\left\langle {2 1 0 _ {   { H } } \rangle}\right.}S U(3)_C \times S U(2)_L \times S U(2)_R \times U(1)_{B-L} \times\left.\mathcal{D}\right|_{M_I} \xrightarrow{\left\langle\overline{{1 2 6}}_{ {H}}\right\rangle} \mathrm{SM}$ \\ [1.5ex]
  \hline
 \end{tabular}
 \caption{Cases of consecutive breakings of $SO(10)$ to SM.}
\label{table:1}
 \end{center}
 \end{table}

Considering the branching rules of:
\begin{equation}
\begin{aligned}
S O(12) & \supset S O(10) \times[U(1)]_{\text {global }} \\
 {1 2} & =( {1})(2)+( {1})(-2)+( {1 0})(0) \\
 {6 6} & =( {1})(0)+( {1 0})(2)+( {1 0})(-2)+( {4 5})(0) \\
 {7 7} & =( {1})(4)+( {1})(0)+( {1})(-4)+( {1 0})(2)+( {1 0})(-2)+( {5 4})(0) \\
 {4 9 5} & =( {4 5})(0)+( {1 2 0})(2)+( {1 2 0})(-2)+( {2 1 0})(0) \\
 {7 9 2} & =( {1 2 0})(0)+( {1 2 6})(0)+(\overline{ {1 2 6}})(0)+( {2 1 0})(2)+( {2 1 0})(-2),
\end{aligned}
\end{equation}
we choose accommodate the Higgs 10 rep into 12 of $SO(12)$ and the $\overline{126}$ that breaks the intermediate gauge group into 792. Regarding the four different breaking scenaria, 45 will come from 66, 54 from 77 and 210 from 792. Examining the $SO(18)$ following branching rules:
\begin{equation}
\begin{aligned}
SO(18) \supset & SO(6) \times SO(12) \\
 {18}= & ({6}, {1})+( {1}, {12}) \\
 {3060}= & ({15}, {1})+({10}, {12})+(\overline{{10}}, {12})+({15}, {66})+({6}, {220})+({1}, {495}) \\
 {8568}= & ({6}, {1})+({15}, {12})+({10}, {66})+(\overline{{10}}, {66})+\\& \qquad+({15}, {220})+({6}, {495})+ ({1}, {792}),
\end{aligned}
\end{equation}
and taking into account the branching rules of eqs. \eqref{so18} and \eqref{153}, we make the following choices regarding the $SO(12)$ reps: 12 comes from 18 of $SO(18)$, 792 from 8568, 66 from 153, 495 from a 3060 and 77 from 170. 

We would like to add a comment about the case of FG. As it was explained in \cite{roumelioti2407}, when attempting to unify FG with internal interactions, along the lines of Unification of Conformal Gravity with $SO(10)$ \cite{Roumelioti:2024lvn}, the difficulties that in principle one is facing are that fermions should \textbf{(a)} be chiral in order to have a chance to survive in low energies and not receive masses as the Planck scale, \textbf{(b)} appear in a matrix representation, since the constructed FG is a matrix model. Then it was suggested \cite{roumelioti2407} and given that the Majorana condition can be imposed, a solution satisfying the conditions \textbf{(a)} and \textbf{(b)} above is the following. We choose to start with $SO(6) \times SO(12)$ as the initial gauge theory with fermions in the $(4, \overline{32})$ representation satisfying in this way the criteria to obtain chiral fermions in tensorial representation of a fuzzy space. Another important point is that using the gauge-theoretic formulation of gravity to construct the FG model, one is led to gauge the $SO(6) \times U(1) \sim SO(2,4) \times U(1)$. Therefore, from this point of view, there only exists a small difference in comparison to the CG case.

\section{Conclusions}

In a previous paper \cite{Roumelioti:2024lvn}, a potentially realistic model was constructed based on the idea that the unification of gravity and internal interactions in four dimensions can be achieved by gauging an enlarged tangent Lorentz group. This possibility was based on the observation that the dimension of the tangent space is not necessarily equal to the dimension of the corresponding curved manifold. In \cite{Roumelioti:2024lvn}, due to the very interesting fact that gravitational theories can be described by gauge theories, CG was firstly constructed in a gauge theoretic manner by gauging the $SO(2,4)$ group. Of particular interest was the fact that the spontaneous symmetry breaking of the constructed CG could lead, among others, to the EG and the WG. Then it is was possible to unify the CG with internal interactions based on the $SO(10)$ GUT (after breaking of $SO(12)$), using the higher-dimensional tangent group $SO(2,16)$. Including fermions and suitably applying the Weyl and Majorana conditions led to a fully unified scheme, which was further examined concerning its behaviour in low energies in ref \cite{Patellis:2024znm}. A similar analysis was done for the FG case, in ref \cite{roumelioti2407}, starting from the $SO(2,4)\times SO(12)$, with fermions in the $(4, \bar{32})$.

\section*{Acknowledgements}
It is a pleasure to thank Thanassis Chatzistavrakidis, Alex Kehagias, Tom Kephart, Spyros Konitopoulos, George Manolakos, Pantelis Manousselis, Carmelo Martin, Tomas Ortin, Gregory Patellis, Roberto Percacci, Manos Saridakis and Nicholas Tracas, for our discussions on various stages of development of the theories presented in the this work.

DR would like to thank NTUA for a fellowship for doctoral studies. GZ would like to thank the Arnold Sommerfeld Centre - LMU Munich for their hospitality and support, the University of Hamburg and DESY for their hospitality,, and the CLUSTER of Excellence 'Quantum Universe', a project of the University of Hamburg in cooperation with DESY for support. 

\printbibliography

@article{Chamseddine2010,
    author = "Chamseddine, A. H. and Mukhanov, Viatcheslav",
    title = "{{G}ravity with de {S}itter and {U}nitary {T}angent {G}roups}",
    eprint = "1002.0541",
    archivePrefix = "arXiv",
    doi = "10.1007/JHEP03(2010)033",
    journal = "JHEP",
    volume = "03",
    pages = "033",
    year = "2010"
}

@article{Chamseddine2016,
    author = "Chamseddine, A. H. and Mukhanov, Viatcheslav",
    title = "{On {U}nification of {G}ravity and {G}auge {I}nteractions}",
    eprint = "1602.02295",
    archivePrefix = "arXiv",
    
    doi = "10.1007/JHEP03(2016)020",
    journal = "JHEP",
    volume = "03",
    pages = "020",
    year = "2016"
}

@article{Heckman_2015,
   title={Covariant non-commutative space–time},
   volume={894},
   url={http://dx.doi.org/10.1016/j.nuclphysb.2015.02.018},
   DOI={10.1016/j.nuclphysb.2015.02.018},
   journal={Nucl. Phys. B},
   publisher={Elsevier BV},
   author={Heckman, J. J. and Verlinde, H.},
   year={2015},
   pages={58–74} 
}

@article{Madore_1992,
doi = {10.1088/0264-9381/9/1/008},
url = {https://dx.doi.org/10.1088/0264-9381/9/1/008},
year = {1992},
publisher = {},
volume = {9},
number = {1},
pages = {69},
author = {J. Madore},
title = {The fuzzy sphere},
journal = {Classical and Quantum Gravity}
}

@article{kastrup_1966,
  title = {Position Operators, Gauge Transformations, and the Conformal Group},
  author = {Kastrup, H. A.},
  journal = {Phys. Rev.},
  volume = {143},
  pages = {1021},
  numpages = {0},
  year = {1966},
  publisher = {APS},
  doi = {10.1103/PhysRev.143.1021},
  url = {https://link.aps.org/doi/10.1103/PhysRev.143.1021}
}

@article{yang1947,
  title = {On Quantized Space-Time},
  author = {Yang, C. N.},
  journal = {Phys. Rev.},
  volume = {72},
  pages = {874},
  numpages = {0},
  year = {1947},
  publisher = {APS},
  doi = {10.1103/PhysRev.72.874},
  url = {https://link.aps.org/doi/10.1103/PhysRev.72.874}
}

@article{Snyder:1946qz,
  title = {Quantized Space-Time},
  author = {Snyder, H. S.},
  journal = {Phys. Rev.},
  volume = {71},
  pages = {38},
  numpages = {0},
  year = {1947},
  publisher = {APS},
  doi = {10.1103/PhysRev.71.38},
  url = {https://link.aps.org/doi/10.1103/PhysRev.71.38}
}

@phdthesis{manolakosphd,
    author = "Manolakos, G.",
    title = "{Construction of gravitational models as noncommutative gauge theories}",
    school = "Natl. Tech. U., Athens",
    year = "2019"
}

@article{Witten:1983,
    author = "Witten, E.",
    journal = "Conf. Proc. C",
    volume = "8306011",
    year = "1983",
pages = "227"
}

@article{Kaluza:1921,
    author = "Kaluza, T.",
    journal = "Sitzungsber. Preuss. Akad. Wiss. Berlin (Math. Phys.)",
pages = "966",
    year = "1921"
}

@article{Li:1973mq,
    author = "Li, L.-F.",
    title = "{Group Theory of the Spontaneously Broken Gauge Symmetries}",
    reportNumber = "SLAC-PUB-1311",
    doi = "10.1103/PhysRevD.9.1723",
    journal = "Phys. Rev. D",
    volume = "9",
    pages = "1723",
    year = "1974"
}

@BOOK{Green2012-ul,
  author = "Green, M.B. and Schwarz, J.H. and Witten, E.",
title = "Superstring Theory",
volume = "1 \& 2",
  year =  "1988"
}

@article{Klein:1926,
    author = "Klein, O.",
    journal = "Z. Phys.",
    volume = "37",
pages = "895",
    year = "1926"
}

@article{Kerner:1968,
    author = "Kerner, R.",
    journal = "Ann. Inst. H. Poincare Phys. Theor.",
    volume = "9",
pages = "143",
    year = "1968"
}

@article{Manolakos:2022universe,
    author = "Manolakos, G. and Manousselis, P. and Roumelioti, D. and Stefas, S. and Zoupanos, G.",
    title = "{A Matrix Model of Four-Dimensional Noncommutative Gravity}",
    journal = "Universe",
    volume = "8",
    number = "4",
    pages = "215",
    year = "2022",
    doi="10.3390/universe8040215"
}

@article{GROSS1985253,
journal = {Nucl. Phys. B},
volume = {256},
pages = "253",
year = {1985},
author = {D.J. Gross and J.A. Harvey and E. Martinec and R. Rohm}
}

@article{CHO1987358,
journal = {J. Math. Phys.},
volume = {16},
year = {1975},
author = {Y.M. Cho},
pages = "2029"
}

@article{PhysRevD.12.1711,
  author = {Cho, Y.M. and Freund, P.G.O.},
  journal = {Phys. Rev. D},
  volume = {12},
  year = {1975},
pages = "1711"
}

@article{FRITZSCH1975193,
journal = {Ann. Phys.},
volume = {93},
year = {1975},
author = {H. Fritzsch and P. Minkowski},
pages= "193"
}

@book{polchinski_1998, 
title = "{String theory}",
volume = "{1 \& 2}",
publisher={Cambridge University Press},
author={Polchinski, J.}, year={1998}
}

@article{forgacs,
author = {P. Forg{\aa}cs and N.S. Manton},
volume = {72},
journal = {Commun. Math. Phys.},
pages = "15",
year = {1980}
}

@article{KAPETANAKIS19924,
journal = {Phys. Rept.},
volume = {219},
author = {D. Kapetanakis and G. Zoupanos},
year = {1992},
pages = "4"
}

@book{Kubyshin:1989vd,
    author = "Kubyshin, Y. A. and Volobuev, I. P. and Mourao, J. M. and Rudolph, G.",
title = "{Dimensional Reduction of Gauge Theories, Spontaneous Compactification and Model Building}",
    publisher = "Springer",
    volume = "349",
    year = "1989"
}

@article{SCHERK197961,
journal = {Nucl. Phys. B},
volume = {153},
year = {1979},
author = {J. Scherk and J.H. Schwarz},
pages = "61"
}

@article{MANTON1981502,
journal = {Nucl. Phys. B},
volume = {193},
year = {1981},
author = {N.S. Manton},
pages = "502"
}

@article{CHAPLINE1982461,
journal = {Nucl. Phys. B},
volume = {209},
year = {1982},
author = {G. Chapline and R. Slansky},
pages = "461"
}

@article{LUST1985309,
journal = {Phys. Lett. B},
volume = {165},
year = {1985},
author = {D. Lust and G. Zoupanos},
pages = "309"
}

@article{Manousselis_2004,
    author = "Manousselis, P. and Zoupanos, G.",
    title = "{Dimensional reduction of ten-dimensional supersymmetric gauge theories in the N=1, D=4 superfield formalism}",
    eprint = "hep-ph/0406207",
    archivePrefix = "arXiv",
    doi = "10.1088/1126-6708/2004/11/025",
    journal = "JHEP",
    volume = "11",
    pages = "025",
    year = "2004"
}

@article{Chatzistavrakidis:2009mh,
    author = "A. Chatzistavrakidis and G. Zoupanos",
    journal = "JHEP",
    volume = "09",
    year = "2009",
pages = "077"
}

@article{utiyama,
  title = {Invariant Theoretical Interpretation of Interaction},
  author = {Utiyama, R.},
  journal = {Phys. Rev.},
  volume = {101},
  pages = {1597},
  numpages = {0},
  year = {1956},
  publisher = {APS},
  doi = {10.1103/PhysRev.101.1597},
  url = {https://link.aps.org/doi/10.1103/PhysRev.101.1597}
}

@article{kibble1961,
    author = {Kibble, T. W. B.},
    title = "{Lorentz Invariance and the Gravitational Field}",
    journal = {Journal of Mathematical Physics},
    number = {2},
    pages = {212-221},
    year = {1961},
    doi = {10.1063/1.1703702},
    url = {https://doi.org/10.1063/1.1703702},
    eprint = {https://pubs.aip.org/aip/jmp/article-pdf/2/2/212/19142021/212\_1\_online.pdf}
}

@article{macdowell,
  title = {Unified Geometric Theory of Gravity and Supergravity},
  author = {MacDowell, S. W. and Mansouri, F.},
  journal = {Phys. Rev. Lett.},
  volume = {38},
  pages = {739},
  year = {1977},
  publisher = {APS},
  doi = {10.1103/PhysRevLett.38.739},
  url = {https://link.aps.org/doi/10.1103/PhysRevLett.38.739}
}

@article{Konitopoulos:2023wst,
    author = "Konitopoulos, S. and Roumelioti, D. and Zoupanos, G.",
    title = "{Unification of Gravity and Internal Interactions}",
    eprint = "2309.15892",
    archivePrefix = "arXiv",
    
    doi = "10.1002/prop.202300226",
    journal = "Fortsch. Phys.",
    volume = "2023",
    pages = "2300226",   year = "2023"
}

@article{Kaku:1978nz,
    author = "Kaku, M. and Townsend, P.K. and van Nieuwenhuizen, P.",
    title = "{Properties of Conformal Supergravity}",
    journal = "Phys. Rev. D",
    volume = "17",
    pages = "3179",
    year = "1978"
}

@article{stellewest,
  title = {Spontaneously broken de Sitter symmetry and the gravitational holonomy group},
  author = {Stelle, K. S. and West, P. C.},
  journal = {Phys. Rev. D},
  volume = {21},
  issue = {6},
  pages = {1466},
  year = {1980},
  publisher = {American Physical Society},
  doi = {10.1103/PhysRevD.21.1466},
  url = {https://link.aps.org/doi/10.1103/PhysRevD.21.1466}
}

@inproceedings{Ivanov:1980tw,
    author = "Ivanov, E.A. and Niederle, J.",
    booktitle = "{9th International Colloquium on Group Theoretical Methods in Physics}",
    year = "1980"
}

@article{KAKU1977304,
title = {Gauge theory of the conformal and superconformal group},
journal = {Physics Letters B},
volume = {69},
number = {3},
pages = {304-308},
year = {1977},
doi = {https://doi.org/10.1016/0370-2693(77)90552-4},
author = {M. Kaku and P.K. Townsend and P. {Van Nieuwenhuizen}}
}

@article{Ivanov:1981wm,
    author = "Ivanov, E.A. and Niederle, J.",
    title = "{Gauge Formulation of Gravitation Theories. 2. The Special Conformal Case}",
    reportNumber = "FZU-81-4",
    doi = "10.1103/PhysRevD.25.988",
    journal = "Phys. Rev. D",
    volume = "25",
    pages = "988",
    year = "1982"
}

@article{Maldacena:2011mk,
    author = "Maldacena, J.",
    title = "{Einstein Gravity from Conformal Gravity}",
    eprint = "1105.5632",
    archivePrefix = "arXiv",
    
    year = "2011",
    note = "e-Print: 1105.5632 [hep-th]"
}

@article{Roumelioti:2024lvn,
    author = "{D. Roumelioti, S. Stefas and G. Zoupanos}",
    title = "{Unification of conformal gravity and internal interactions}",
    eprint = "2403.17511",
    archivePrefix = "arXiv",
    
    doi = "10.1140/epjc/s10052-024-12949-6",
    journal = "Eur. Phys. J. C",
    volume = "84",
    number = "6",
    pages = "577",
    year = "2024"
}

@article{Hell:2023rbf,
    author = "Hell, A. and Lust, D. and Zoupanos, G.",
    title = "{On the ghost problem of conformal gravity}",
    eprint = "2306.13714",
    archivePrefix = "arXiv",
    
    reportNumber = "CERN-TH-2023-116, LMU-ASC 21/23, MPP-2023-134",
    doi = "10.1007/JHEP08(2023)168",
    journal = "JHEP",
    volume = "08",
    pages = "168",
    year = "2023"
}

@article{Anastasiou:2016jix,
    author = "Anastasiou, G. and Olea, R.",
    title = "{From conformal to Einstein Gravity}",
    eprint = "1608.07826",
    archivePrefix = "arXiv",
    
    doi = "10.1103/PhysRevD.94.086008",
    journal = "Phys. Rev. D",
    volume = "94",
    number = "8",
    pages = "086008",
    year = "2016"
}

@article{Fradkin:1985am,
    author = "Fradkin, E.S. and Tseytlin, A.A.",
    title = "{Conformal Supergravity}",
    doi = "10.1016/0370-1573(85)90138-3",
    journal = "Phys. Rept.",
    volume = "119",
    pages = "233",
    year = "1985"
}

@article{Manolakos_paper1,
   title={Four-dimensional gravity on a covariant noncommutative space},
   volume={2020},
   number={8},
   journal={JHEP},
   publisher={Springer Science and Business Media LLC},
   author={Manolakos, G. and Manousselis, P. and Zoupanos, G.},
   year={2020},
url="
https://doi.org10.1007/jhep08(2020)001"
}

@article{Manolakos_paper2,
author = {Manolakos, G. and Manousselis, P. and Zoupanos, G.},
title = {Four-Dimensional Gravity on a Covariant Noncommutative Space ({II})},
journal = {Fortsch. Phys.},
volume = {69},
number = {8-9},
pages = {2100085},
doi = {https://doi.org/10.1002/prop.202100085},
url ={https://onlinelibrary.wiley.com/doi/abs/10.1002/prop.202100085},
year = {2021}
}

@article{Manolakos:2023hif,
    author = "Manolakos, G. and Manousselis, P. and Roumelioti, D. and Stefas, S. and Zoupanos, G.",
    title = "{Intertwining noncommutativity with gravity and particle physics}",
    eprint = "2305.11785",
    archivePrefix = "arXiv",
    
    doi = "10.1140/epjs/s11734-023-00830-8",
    journal = "Eur. Phys. J. ST",
    volume = "232",
    number = "23-24",
    pages = "3607",
    year = "2023"
}

@article{Kibble:1985sn,
    author = "Kibble, T.W.B. and Stelle, K.S.",
    year = "1985",
    title = "{Gauge theories of gravity and supergravity}",
journal = "{Progress In Quantum Field
Theory}"   
}

@book{freedman_vanproeyen_2012,
    author = "Freedman, Daniel Z. and Van Proeyen, Antoine",
    title = "{Supergravity}",
    doi = "10.1017/CBO9781139026833",
    publisher = "Cambridge Univ. Press",
    year = "2012"
}

@article{Chatzistavrakidis_2018,
year = 2018,
author = "Chatzistavrakidis, A. and Jonke, L. and Jurman, D. and Manolakos, G. and Manousselis, P. and Zoupanos, G.",
volume = {66},
journal = {Fortsch. Phys.},
    pages = "1800047"
}

@article{Weinberg:1984ke,
    author = "Weinberg, Steven",
    title = "{Generalized Theories of Gravity and Supergravity in Higher Dimensions}",
    journal = "{Fifth Workshop on Grand Unification}",
    reportNumber = "UTTG-12-84",
    year = "1984",
    url="https://lib-extopc.kek.jp/preprints/PDF/1984/8408/8408083.pdf"
}

@article{Percacci:1984ai,
    author = "Percacci, R.",
    title = "{Spontaneous Soldering}",
    doi = "10.1016/0370-2693(84)90171-0",
    journal = "Phys. Lett. B",
    volume = "144",
    pages = "37",
    year = "1984"
}

@article{Percacci_1991,  
    author = "Percacci, R.",
    title = "{The Higgs phenomenon in quantum gravity}",
    eprint = "0712.3545",
    archivePrefix = "arXiv",
    reportNumber = "SISSA-106-90-EP",
    doi = "10.1016/0550-3213(91)90510-5",
    journal = "Nucl. Phys. B",
    volume = "353",
    pages = "271",
    year = "1991"
}

@article{Nesti_2008,
    author = "Nesti, Fabrizio and Percacci, Roberto",
    title = "{Graviweak Unification}",
    eprint = "0706.3307",
    archivePrefix = "arXiv",
    doi = "10.1088/1751-8113/41/7/075405",
    journal = "J. Phys. A",
    volume = "41",
    pages = "075405",
    year = "2008"
}

@article{Nesti_2010,
    author = "Nesti, F. and Percacci, R.",
    title = "{Chirality in unified theories of gravity}",
    eprint = "0909.4537",
    archivePrefix = "arXiv",
    doi = "10.1103/PhysRevD.81.025010",
    journal = "Phys. Rev. D",
    volume = "81",
    pages = "025010",
    year = "2010"
}

@article{Krasnov:2017epi,
    author = "Krasnov, Kirill and Percacci, Roberto",
    title = "{Gravity and Unification: A review}",
    eprint = "1712.03061",
    archivePrefix = "arXiv",
    doi = "10.1088/1361-6382/aac58d",
    journal = "Class. Quant. Grav.",
    volume = "35",
    number = "14",
    pages = "143001",
    year = "2018"
}

@article{D_Auria_2001,
    author = "D'Auria, R. and Ferrara, S. and Lledo, M. A. and Varadarajan, V. S.",
    title = "{Spinor algebras}",
    eprint = "hep-th/0010124",
    archivePrefix = "arXiv",
    reportNumber = "CERN-TH-2000-260",
    doi = "10.1016/S0393-0440(01)00023-7",
    journal = "J. Geom. Phys.",
    volume = "40",
    pages = "101",
    year = "2001"
}

@article{noncomtomos,
author = {Schupp, P. and Anagnostopoulos, K. and Zoupanos, G.},
year = {2024},
pages = {1},
title = {Noncommutativity and physics},
volume = {232},
journal = {Eur.Phys.J.ST},
doi = {10.1140/epjs/s11734-024-01086-6}
}

@article{majoranaspinors,
author = {Figueroa - O’Farrill, J.},
title = {Lecture notes on Majorana Spinors}, 
journal = {School of Mathematics, Universiry of Edinburgh},
year = {www.maths.ed.ac.uk/~jmf/Teaching/Lectures/Majorana.pdf}
}

@article{Chamseddine:2002fd,
    author = "Chamseddine, A. H.",
    title = "{Invariant actions for noncommutative gravity}",
    eprint = "hep-th/0202137",
    archivePrefix = "arXiv",
    reportNumber = "CAMS-02-02",
    doi = "10.1063/1.1572199",
    journal = "J. Math. Phys.",
    volume = "44",
    pages = "2534",
    year = "2003"
}

@article{Feger_2020,
	year = 2020,
	volume = {257},
	author = {R. Feger and T.W. Kephart and R.J. Saskowski},
	journal = {Comput. Phys. Commun.},
        pages = {107490}
}

@article{Slansky:1981yr,
    author = "Slansky, R.",
    journal = "Phys. Rept.",
    volume = "79",
    year = "1981",
    pages = {1}
}

@BOOK{Georgi1999,
  author    = "Georgi, H.",
title = "{L}ie {A}lgebras {I}n {P}article {P}hysics: from {I}sospin To {U}nified {T}heories",
  series    = "Frontiers in Physics",
  year      =  "1999"
}

@book{Lust:1989tj,
    author = "Lust, D. and Theisen, S.",
    title = "{Lectures on {S}tring {T}heory}",
    volume = "346",
    year = "1989"
}

@article{Irges:2011de,
    author = "Irges, Nikos and Zoupanos, George",
    eprint = "1102.2220",
    archivePrefix = "arXiv",
    primaryClass = "hep-ph",
    doi = "10.1016/j.physletb.2011.03.005",
    journal = "Phys. Lett. B",
    volume = "698",
    pages = "146",
    year = "2011"
}

@article{Manolakos:2020cco,
    author = "Manolakos, George and Patellis, Gregory and Zoupanos, George",
    eprint = "2009.07059",
    archivePrefix = "arXiv",
    primaryClass = "hep-ph",
    doi = "10.1016/j.physletb.2020.136031",
    journal = "Phys. Lett. B",
    volume = "813",
    pages = "136031",
    year = "2021"
}

@article{Patellis:2024dfl,
    author = "Patellis, Gregory and Porod, Werner and Zoupanos, George",
    title = "{Split NMSSM from dimensional reduction of a $10D$, $\mathcal{N}=1$, $E_8$ theory over a modified flag manifold}",
    eprint = "2307.10014 [hep-ph]",
    journal = "JHEP",
    volume = "1",
    pages = "021",
    year = "2024",
    doi="10.48550/arXiv.2307.10014"
}

@article{Sciama,
    author = "Sciama, D.W.",
    title = "{Recent developments in the theory of gravitational radiation}",
    journal = "Gen Relat Gravit",
    volume = "3",
    pages = "149--165",
    year = "1972",
    doi="10.1007/BF00755934"
}

@article{Umezawa,
    author = "Umezawa, H.",
    title = "{Dynamical rearrangement of symmetries}",
    journal = "Nuovo Cimento A (1965-1970)",
    volume = "40",
    pages = "450--475",
    year = "1965",
    doi="10.1007/BF02721035"
}

@article{Matsumoto,
  title = {Infrared effect of the Goldstone boson and the order parameter},
  author = {Matsumoto, H. and Papastamatiou, N. J. and Umezawa, H.},
  journal = {Phys. Rev. D},
  volume = {12},
  issue = {6},
  pages = {1836--1839},
  numpages = {0},
  year = {1975},
  doi = {10.1103/PhysRevD.12.1836},
  url = {10.1103/PhysRevD.12.1836}
}

@book{Ortín_2015,
    series={Cambridge Monographs on Mathematical Physics}, 
    title={Gravity and Strings}, 
    author={Ortín, Tomás}, 
    year={2015}, 
    collection={Cambridge Monographs on Mathematical Physics},
    doi = {10.1017/CBO9781139019750}
}

@article{castellani,
  title = {$OS_p(1|4)$ Supergravity and its noncommutative extension},
   volume={88},
   ISSN={1550-2368},
   url={http://dx.doi.org/10.1103/PhysRevD.88.025022},
   doi={10.1103/physrevd.88.025022},
   number={2},
   journal={Phy. Rev. D},
   author={Castellani, Leonardo},
   year={2013}
}

@article{weyl,
    author = {Weyl, H.},
    title = {Gravitation and the electron},
    journal = {PNAS},
    volume = {15},
    number = {4},
    pages = {323-334},
    year = {1929},
    doi = {10.1073/pnas.15.4.323},
    URL = {https://www.pnas.org/doi/abs/10.1073/pnas.15.4.323},
    eprint = {https://www.pnas.org/doi/pdf/10.1073/pnas.15.4.323}
}

@article{mannheim,
  title = {{Making the Case for Conformal Gravity}},
   volume={42},
   ISSN={388--420},
   doi={10.1007/s10701-011-9608-6},
   journal={Found. Phys.},
   author={Mannheim, P.D},
   year={2012}
}

@article{roumelioti2407,
author = {Roumelioti, Danai and Stefas, Stelios and Zoupanos, George},
title = {{Fuzzy Gravity: Four-Dimensional Gravity on a Covariant Noncommutative Space and Unification with Internal Interactions}},
journal = {Fortschr. Phys.},
volume = {72},
number = {9-10},
pages = {2400126},
doi = {https://doi.org/10.1002/prop.202400126},
url = {https://onlinelibrary.wiley.com/doi/abs/10.1002/prop.202400126},
eprint = {https://onlinelibrary.wiley.com/doi/pdf/10.1002/prop.202400126},
year = {2024}
}

@article{ghilencea2023,
  title = {{Weyl conformal geometry vs Weyl anomaly}},
   ISSN={113},
   doi={10.1007/JHEP10(2023)113},
   journal={JHEP},
   author={Ghilencea, D.M.},
   year={2023},
volume = "10",
    pages = "113"
}

@article{weyl1929,
  title = {{Elektron und Gravitation}},
   volume={56},
   pages={330--352},
   doi={10.1007/BF01339504},
   journal={I. Z. Physik},
   author={Weyl, H.},
   year={1929}
}

@article{Patellis:2024znm,
    author = "Patellis, Gregory and Tracas, Nicholas and Zoupanos, George",
    title = "{From the Unification of Conformal and Fuzzy Gravities with Internal Interactions to the $SO(10)$ GUT and the Particle Physics Standard Model}",
    note = "e-Print: 2412.02786, to be published in Eur. Phys. J. C",
    archivePrefix = "arXiv",
    reportNumber = "IFT-UAM/CSIC-24-159",
    year = "2024"
}

@article{Manousselis:2001re,
    author = "Manousselis, P. and Zoupanos, G.",
    title = "{Dimensional reduction over coset spaces and supersymmetry breaking}",
    eprint = "hep-ph/0111125",
    archivePrefix = "arXiv",
    reportNumber = "NTUA-08-01",
    doi = "10.1088/1126-6708/2002/03/002",
    journal = "JHEP",
    volume = "03",
    pages = "002",
    year = "2002"
}

@article{Manousselis:2000aj,
    author = "Manousselis, P. and Zoupanos, G.",
    title = "{Supersymmetry breaking by dimensional reduction over coset spaces}",
    eprint = "hep-ph/0010141",
    archivePrefix = "arXiv",
    reportNumber = "NTUA-96-00",
    doi = "10.1016/S0370-2693(01)00268-4",
    journal = "Phys. Lett. B",
    volume = "504",
    pages = "122--130",
    year = "2001"
}

@article{Manousselis:2001xb,
    author = "Manousselis, P. and Zoupanos, G.",
    title = "{Soft supersymmetry breaking due to dimensional reduction over nonsymmetric coset spaces}",
    eprint = "hep-ph/0106033",
    archivePrefix = "arXiv",
    reportNumber = "NTUA-05-01",
    doi = "10.1016/S0370-2693(01)01040-1",
    journal = "Phys. Lett. B",
    volume = "518",
    pages = "171--180",
    year = "2001"
}

@article{Condeescu:2023izl,
    author = "Condeescu, C. and Ghilencea, D. M. and Micu, A.",
    title = "{Weyl quadratic gravity as a gauge theory and non-metricity vs torsion duality}",
    eprint = "2312.13384",
    archivePrefix = "arXiv",
    primaryClass = "hep-th",
    doi = "10.1140/epjc/s10052-024-12644-6",
    journal = "Eur. Phys. J. C",
    volume = "84",
    number = "3",
    pages = "292",
    year = "2024"
}
\end{document}